\documentclass[preprint,aps,showpacs,floatfix]{revtex4}
\usepackage{epsfig}
\usepackage{bm}

\def\e{\kern+.5ex\lower.42ex\hbox{$\scriptstyle \iota$}\kern-1.10ex e}

\newcommand{\beq}{\begin{equation}}
\newcommand{\eeq}{\end{equation}}
\newcommand{\beqa}{\begin{eqnarray}}
\newcommand{\eeqa}{\end{eqnarray}}
\newcommand{\nn}{\nonumber \\ }
\newcommand{\fet}[1]{\mbox{\boldmath $#1$}}

\begin{document}

\title{
A first estimation of chiral four-nucleon force effects in \bm{$^4$}He
}

\author{D. Rozp{\e}dzik, J.~Golak, R.~Skibi\'nski, H.~Wita{\l}a}
\affiliation{M. Smoluchowski Institute of Physics, Jagiellonian University,
                    PL-30059 Krak\'ow, Poland}

\author{W.~Gl\"ockle}
\affiliation{Institut f\"ur Theoretische Physik II,
         Ruhr Universit\"at Bochum, D-44780 Bochum, Germany}

\author{E. Epelbaum}
\affiliation{Forschungszentrum J\"ulich, IKP
(Theorie), D-52425 J\"ulich, Germany}
\affiliation{Helmholtz-Institut f\"ur Strahlen- und Kernphysik (Theorie), Universit\"at
Bonn, Nu{\ss}allee 14-16, D-53115 Bonn, Germany}

\author{A.~Nogga}
\affiliation{Forschungszentrum J\"ulich, IKP
(Theorie), D-52425 J\"ulich, Germany}

\author{H.~Kamada}
\affiliation{Department of Physics, Faculty of Engineering,
  Kyushu Institute of Technology,
  1-1 Sensuicho, Tobata, Kitakyushu 804-8550, Japan}

\date{\today}

\begin{abstract}
We estimate four-nucleon force effects
between different $^4$He wave functions 
by calculating the expectation values
of four--nucleon potentials
which were recently derived
within the framework of chiral effective field theory.
We find that the four-nucleon force is attractive 
for the wave functions with a totally symmetric momentum part.
The additional binding energy provided by the long-ranged part 
of the four-nucleon force is of the order of a few hundred keV.
\end{abstract}


\maketitle



\section{Introduction}
\label{sec:1}
\begin{figure}[tb]
\includegraphics[width=8.5cm,keepaspectratio,angle=0,clip]{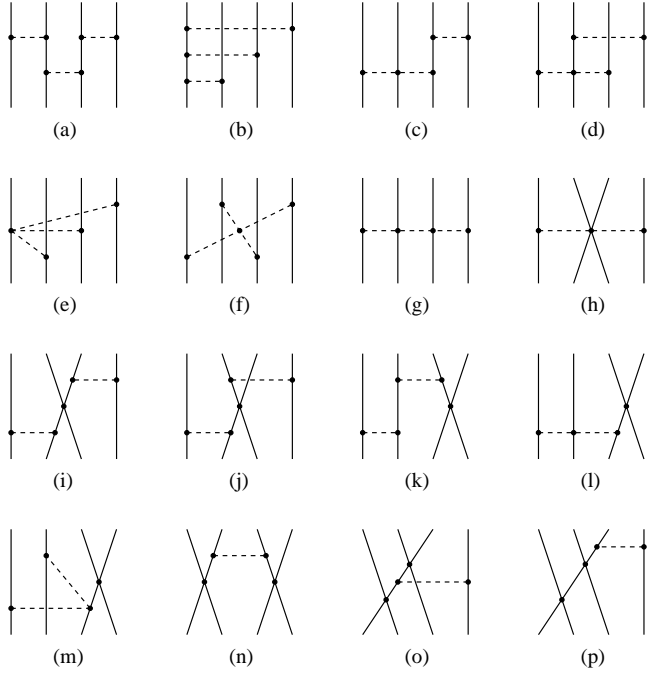}
\caption{
The leading contributions to the four--nucleon force.
Solid and dashed lines represent nucleons and pions,
respectively. Graphs resulting by the interchange of the 
vertex ordering and/or nucleon lines are not shown.
\label{fig1}
}
\end{figure}

In a recent paper \cite{4nfpaper} the leading contribution to the four--nucleon force, $V_{4N}$,
has been derived within the framework of chiral effective field theory.
It is governed by the exchange of pions and the lowest--order
nucleon--nucleon contact interaction and includes effects due to the nonlinear
pion--nucleon couplings and the pion self interactions constrained by the chiral
symmetry of QCD.
The individual pieces of $V_{4N}$ corresponding to the diagrams 
in Fig.~\ref{fig1} read \cite{4nfpaper}
\beqa
\label{4nf}
V^{a} &=& - \frac{2 g_A^6}{( 2 F_\pi )^6}
\frac{\vec \sigma_1 \cdot \vec q_1 \;\vec \sigma_4 \cdot \vec q_4}{[\vec q_1^{\;2}  + M_\pi^2]\,
[\vec q_{12}^{\;2}  + M_\pi^2]^2 \, [\vec q_4^{\;2}  + M_\pi^2]} \nn
&\times& \Big[ ( \fet \tau_1 \cdot \fet \tau_4 \,  \fet \tau_2 \cdot \fet \tau_3
-  \fet \tau_1 \cdot \fet \tau_3 \,  \fet \tau_2 \cdot \fet \tau_4 ) \,\vec q_1 \cdot \vec q_{12}
\, \vec q_4 \cdot \vec q_{12} \nn
&& {} + \fet \tau_1 \times \fet \tau_2 \cdot \fet \tau_4  \; \vec q_1 \cdot \vec q_{12}  \;
\vec q_{12} \times \vec q_4 \cdot \vec \sigma_3 \nn
&& {}  + \fet \tau_1 \times \fet \tau_3 \cdot \fet \tau_4  \; \vec q_4 \cdot \vec q_{12}  \;
\vec q_{1} \times \vec q_{12} \cdot \vec \sigma_2  \nn
&& {} + \fet \tau_1 \cdot \fet \tau_4 \; \vec q_{12} \times \vec q_{1} \cdot \vec \sigma_2 \;
\vec q_{12} \times \vec q_4 \cdot \vec \sigma_3  \Big] +\mbox{all permutations},\nn [4pt]
V^{c} &=& - \frac{2 g_A^4}{(2 F_\pi)^6}
\frac{\vec \sigma_1 \cdot \vec q_1 \;\vec \sigma_4 \cdot \vec q_4}{[\vec q_1^{\; 2}  + M_\pi^2]\,
[\vec q_{12}^{\; 2}  + M_\pi^2]\, [\vec q_4^{\; 2}  + M_\pi^2]}  \nn
&& {} \times \Big[ (  \fet \tau_1 \cdot \fet \tau_4 \,  \fet \tau_2 \cdot \fet \tau_3
-  \fet \tau_1 \cdot \fet \tau_3 \,  \fet \tau_2 \cdot \fet \tau_4 ) \, \vec q_{12} \cdot \vec q_4 \nn
&& {} +  \fet \tau_1 \times \fet \tau_2 \cdot \fet \tau_4  \; \vec q_{12} \times \vec q_4 \cdot \vec \sigma_3 \Big] +\mbox{all permutations}, \nn [4pt]
V^{e} &=& \frac{g_A^4}{(2 F_\pi)^6}
\frac{\vec \sigma_2 \cdot \vec q_2 \;\vec \sigma_3 \cdot \vec q_3 \;\vec \sigma_4 \cdot \vec q_4}{[\vec q_2^{\;2}  + M_\pi^2]\,
[\vec q_{3}^{\;2}  + M_\pi^2] \, [\vec q_4^{\;2}  + M_\pi^2]}  \nn [1pt]
&& {} \times \fet \tau_1 \cdot \fet \tau_2 \,  \fet \tau_3 \cdot \fet \tau_4 \;
\vec \sigma_1 \cdot (\vec q_3 + \vec q_4 ) + \mbox{all permutations}, \nn   [4pt]
V^{f} &=& \frac{g_A^4}{2 (2 F_\pi)^6}
\; \Big[ \left( \vec q_1 + \vec q_2 \, \right)^2 + M_\pi^2 \Big] \nn
&& {} \times
\frac{\vec \sigma_1 \cdot \vec q_1 \;\vec \sigma_2 \cdot \vec q_2 \;\vec \sigma_3 \cdot \vec q_3 \;\vec \sigma_4 \cdot \vec q_4}
{[\vec q_1^{\;2}  + M_\pi^2]\,  [\vec q_{2}^{\;2}  + M_\pi^2] \, [\vec q_{3}^{\;2}  + M_\pi^2] \, [\vec q_4^{\;2}  + M_\pi^2]} \nn  [2pt]
&& {} \times  \fet \tau_1 \cdot \fet \tau_2 \,  \fet \tau_3 \cdot \fet \tau_4   + \mbox{all permutations}, \nn [4pt]
V^{k} &=& 4 C_T \frac{g_A^4}{(2 F_\pi)^4} \,
\frac{\vec \sigma_1 \cdot \vec q_1 \; \vec \sigma_3 \times \vec \sigma_4  \cdot \vec q_{12}}
{[\vec q_1^{\;2}  + M_\pi^2]\, [\vec q_{12}^{\;2}  + M_\pi^2]^2} \nn
&& {} \times  \Big[ \fet \tau_1 \cdot \fet \tau_3 \; \vec q_1 \times \vec q_{12} \cdot \vec \sigma_2   -
\fet \tau_1 \times \fet \tau_2 \cdot \fet \tau_3 \; \vec q_{1} \cdot \vec q_{12} \Big] \nn [1pt]
&& {} +  \mbox{all permutations}, \nn [4pt]
V^l &=& - 2 C_T \frac{g_A^2}{(2 F_\pi)^4} \,
\frac{\vec \sigma_1 \cdot \vec q_1 \; \vec \sigma_3 \times \vec \sigma_4  \cdot \vec q_{12}}
{[\vec q_1^{\;2}  + M_\pi^2] \, [\vec q_{12}^{\;2}  + M_\pi^2]} \; \fet \tau_1 \times \fet \tau_2 \cdot \fet \tau_3   \nn [2pt]
&& {} + \mbox{all permutations}, \nn[4pt]
V^n &=&  2 C_T^2 \frac{g_A^2}{(2 F_\pi)^2} \,
\frac{\vec \sigma_1 \times \vec \sigma_2  \cdot \vec q_{12} \; \vec \sigma_3 \times \vec \sigma_4  \cdot \vec q_{12}}
{[\vec q_{12}^{\;2}  + M_\pi^2]^2} \;  \fet \tau_2 \cdot \fet \tau_3  \nn
&& {} + \mbox{all permutations} .
\eeqa
Here, the subscripts refer to the nucleon labels and $\vec q_{i} = \vec p_i \, ' - \vec p_i$ with $\vec p_i \, '$
and $\vec p_i$ being the final and initial momenta of the nucleon i.
Further, $\vec q_{12} = \vec q_1 + \vec q_2 = - \vec q_3 - \vec q_4 = -\vec q_{34}$ is the momentum transfer between the
nucleon pairs 12 and 34.  Diagrams (b), (d), (g), (h), (i), (j), (m), (o) and (p) lead to vanishing contributions
to the four-nucleon (4N) force. The total short--range 4N force depends only on one low--energy constant $C_T$.

\section{Calculations}
\label{sec:2}

We would like to estimate
the magnitude of that 4N force
in the 4N bound state.
In order to simplify the calculations in a first attempt
we assume that the
momentum part of the $^4$He wave function is totally symmetric
with respect to any permutations of the nucleons.
Thus we deal with the totally antisymmetric spin-isospin part 
$ \mid \xi \rangle $ of the total wave function  
\beqa
\mid \xi \rangle 
= \frac1{\sqrt{2}} \, \left(
\left\{ \mid s_{12}=1 , t_{12}=0 > \mid s_{34}=1 , t_{34}=0 > \right\}_{S=0,T=0}  \right.
\nonumber \\
- \left. \left\{ \mid s_{12}=0 , t_{12}=1 > \mid s_{34}=0 , t_{34}=1 > \right\}_{S=0,T=0}
                      \right) \, ,
\label{xi}
\eeqa
where $s_{ij}$ and $t_{ij}$ are the total two-nucleon subsystem spins and isospins.
The curly brackets denote the coupling of the subsystems spins and isospins
to the total spin ($S=0$) and isospin ($T=0$) of the 4N bound state.
The state $ \mid \xi \rangle $ can be expanded into the sum of product states 
\begin{eqnarray}
\mid \xi \rangle = &  & \nonumber \\
\frac1{\sqrt{24}} \, \left\{  \right. &
-\mid - + - + \rangle \mid - - + + \rangle  
+\mid + - - + \rangle \mid - - + + \rangle 
+\mid - + + - \rangle \mid - - + + \rangle & \nonumber \\ 
& -\mid + - + - \rangle \mid - - + + \rangle 
+\mid - - + + \rangle \mid - + - + \rangle 
-\mid + - - + \rangle \mid - + - + \rangle & \nonumber \\ 
& -\mid - + + - \rangle \mid - + - + \rangle 
+\mid + + - - \rangle \mid - + - + \rangle 
-\mid - - + + \rangle \mid + - - + \rangle & \nonumber \\ 
& +\mid - + - + \rangle \mid + - - + \rangle 
+\mid + - + - \rangle \mid + - - + \rangle  
-\mid + + - - \rangle \mid + - - + \rangle  & \nonumber \\ 
& -\mid - - + + \rangle \mid - + + - \rangle  
+\mid - + - + \rangle \mid - + + - \rangle  
+\mid + - + - \rangle \mid - + + - \rangle  & \nonumber \\ 
& -\mid + + - - \rangle \mid - + + - \rangle  
+\mid - - + + \rangle \mid + - + - \rangle  
-\mid + - - + \rangle \mid + - + - \rangle  & \nonumber \\ 
& -\mid - + + - \rangle \mid + - + - \rangle  
+\mid + + - - \rangle \mid + - + - \rangle  
-\mid - + - + \rangle \mid + + - - \rangle  & \nonumber \\ 
& +\mid + - - + \rangle \mid + + - - \rangle  
+\mid - + + - \rangle \mid + + - - \rangle  
-\mid + - + - \rangle \mid + + - - \rangle  \left. \right\} &  \nonumber \\
\equiv &  \frac1{\sqrt{24}} \sum\limits_{i=1}^{24} \, s(i) \mid
\chi_1(i) \chi_2(i) \chi_3(i) \chi_4(i) \rangle \, \mid \eta_1(i)
\eta_2(i) \eta_3(i) \eta_4(i) \rangle , \ \ \ \ \ \ \ \ \ 
\ \ \ \ \ \ \ \ \ \ \ \ \ \ &
\end{eqnarray}
where $\chi_j(i)$ ($\eta_j(i)$) is the spin (isospin) state of the j$^{\rm th}$ 
nucleon in the i$^{\rm th}$ term of the sum, and $s(i)$ denotes the sign of the 
i$^{\rm th}$ product state.  The ``+'' and ``-'' signs inside the kets 
stand for the $+\frac12$ and $-\frac12$ spin and isospin projections, respectively.
All the single nucleon states are normalized to $1$
\begin{eqnarray}
\langle \chi_j(i) \mid \chi_j(i) \rangle =  
\langle \eta_j(i) \mid \eta_j(i) \rangle = 1 
\label{norm}
\end{eqnarray}
and consequently also the state $\mid \xi \rangle $ has the same norm
\begin{eqnarray}
\langle \xi \mid \xi \rangle = 1  .
\label{normxi}
\end{eqnarray}

The momentum part of the total wave function in the 4N center of mass (c.m.) system
depends on three relative (Jacobi) momenta
\begin{eqnarray}
{\vec p} & = & \frac{ {\vec p}_1 - {\vec p}_2 }{2} \nonumber \\
{\vec q} & = & \frac{ 2{\vec p}_3 - \left({\vec p}_1 + {\vec p}_2 \right) }{3} \nonumber \\
{\vec t} & = & \frac{ 3{\vec p}_4 - \left({\vec p}_1 + {\vec p}_2 + {\vec p}_3 \right) }{4} \, ,
\label{jacobi}
\end{eqnarray}
where ${\vec p}_i$ are the individual nucleon momenta.
Equations (\ref{jacobi}) can be inverted in order to express the individual momenta in terms
of the relative momenta ${\vec p}$, ${\vec q}$ and ${\vec t}$:
\begin{eqnarray}
{\vec p}_1 & = & \frac{ 6{\vec p} - 3{\vec q} - 2{\vec t}}{6} \nonumber \\
{\vec p}_2 & = & \frac{ -6{\vec p} - 3{\vec q} - 2{\vec t}}{6} \nonumber \\
{\vec p}_3 & = & \frac{  3{\vec q} - {\vec t}}{3} \nonumber \\
{\vec p}_4 & = & {\vec t} .
\label{indiv}
\end{eqnarray}

The assumption that the momentum part of the $^4$He wave function is totally symmetric
is still very general and we make further restrictions. We assume 
that the momentum part can be written as a function of one variable, $x$, where
\begin{eqnarray}
x\equiv  \frac1{2m}\, \left( {\vec p}_1^{\ 2} + {\vec p}_2^{\ 2} + {\vec p}_3^{\ 2} + {\vec p}_4^{\ 2}\right)
\, = \,
\frac1{m} \left( {\vec p}^{\ 2} + \frac34 {\vec q}^{\ 2}  + \frac23 {\vec t}^{\ 2} \right) \, ,
\label{x}
\end{eqnarray}
which is the c.m. kinetic energy of the 4N system. ($m$ is the nucleon mass.)
This implicitly means that we set all angular momenta to zero. 
We will later show to what extent this choice is realistic.
Consequently we can write the full wave function $\mid \Psi \rangle $
as
\begin{eqnarray}
\langle {\vec p}\,  {\vec q} \, {\vec t} \mid \Psi \rangle = f (x) \mid \xi \rangle .
\label{wf}
\end{eqnarray}

In order to calculate the matrix elements 
$\langle \Psi \mid V_{4N} \mid \Psi \rangle  $
we calculate first the matrix elements in the spin-isospin space 
for the pieces of the 4N force given in Eq.~(\ref{4nf}).
For $V^a$ we consider first the following expression
\begin{eqnarray}
\langle \xi \mid V^{a}_1 \mid \xi \rangle 
\equiv 
\left( {\vec q}_1 \cdot {\vec q}_{12} \right) \, 
\left( {\vec q}_4 \cdot {\vec q}_{12} \right) \, 
\langle \xi \mid 
\left( {\vec \sigma}_1 \cdot {\vec q}_1 \right) \, 
\left( {\vec \sigma}_4 \cdot {\vec q}_4 \right) \, 
 {\vec \tau}_1 \cdot {\vec \tau}_4 \, 
 {\vec \tau}_2 \cdot {\vec \tau}_3 \, 
 \mid \xi \rangle  \nonumber \\
= \frac1{24} \, 
\left( {\vec q}_1 \cdot {\vec q}_{12} \right) \, 
\left( {\vec q}_4 \cdot {\vec q}_{12} \right) \, 
\sum\limits_{i=1}^{24} \,
\sum\limits_{j=1}^{24} \,
\sum\limits_{\alpha, \beta, \gamma, \delta=1}^3 \,
s(i) s(j) q_1(\alpha) q_4(\beta) 
\langle \chi_1(j) \mid \sigma_\alpha \mid \chi_1(i) \rangle \, \nonumber \\
\langle \chi_2(j) \mid \chi_2(i) \rangle \,
\langle \chi_3(j) \mid \chi_2(i) \rangle \,
\langle \chi_4(j) \mid \sigma_\beta \mid \chi_4(i) \rangle \, \nonumber \\
\langle \eta_1(j) \mid \tau_\gamma \mid \eta_1(i) \rangle \, 
\langle \eta_2(j) \mid \tau_\delta \mid \eta_2(i) \rangle \, 
\langle \eta_3(j) \mid \tau_\delta \mid \eta_3(i) \rangle \, 
\langle \eta_4(j) \mid \tau_\gamma \mid \eta_4(i) \rangle \, \nonumber \\
= 
\left( {\vec q}_1 \cdot {\vec q}_{12} \right) \, 
\left( {\vec q}_4 \cdot {\vec q}_{12} \right) \, 
\left( {\vec q}_1 \cdot {\vec q}_{4} \right) \, .
\label{meVa1}
\end{eqnarray}
The intermediate multiple sums in Eq.~(\ref{meVa1}) 
were obtained by means of the {\em Mathematica} program.
In the same way we obtain the other expressions
\begin{eqnarray}
\langle \xi \mid V^{a}_2 \mid \xi \rangle 
\equiv 
-\left( {\vec q}_1 \cdot {\vec q}_{12} \right) \, 
\left( {\vec q}_4 \cdot {\vec q}_{12} \right) \, 
\langle \xi \mid 
\left( {\vec \sigma}_1 \cdot {\vec q}_1 \right) \, 
\left( {\vec \sigma}_4 \cdot {\vec q}_4 \right) \, 
 {\vec \tau}_1 \cdot {\vec \tau}_3 \, 
 {\vec \tau}_2 \cdot {\vec \tau}_4 \, 
 \mid \xi \rangle  \nonumber \\
=3 \,\left( {\vec q}_1 \cdot {\vec q}_{12} \right) \, 
\left( {\vec q}_4 \cdot {\vec q}_{12} \right) \, 
\left( {\vec q}_1 \cdot {\vec q}_{4} \right) \, ,
\label{meVa2}
\end{eqnarray}
\begin{eqnarray}
\langle \xi \mid V^{a}_3 \mid \xi \rangle 
\equiv 
\left( {\vec q}_1 \cdot {\vec q}_{12} \right) \, 
\langle \xi \mid 
\left( {\vec \sigma}_1 \cdot {\vec q}_1 \right) \, 
\left( {\vec \sigma}_4 \cdot {\vec q}_4 \right) \, 
\left( {\vec q}_{12} \times {\vec q}_4 \right) \cdot {\vec \sigma}_3 \, 
\left( {\vec \tau}_{1} \times {\vec \tau}_2 \right) \cdot {\vec \tau}_4 \, 
 \mid \xi \rangle  \nonumber \\
=-2 \,\left( {\vec q}_1 \cdot {\vec q}_{12} \right) \, 
\left( {\vec q}_{12} \times {\vec q}_{4} \right) \cdot
\left( {\vec q}_1 \times {\vec q}_{4} \right) \, ,
\label{meVa3}
\end{eqnarray}
\begin{eqnarray}
\langle \xi \mid V^{a}_4 \mid \xi \rangle 
\equiv 
\left( {\vec q}_4 \cdot {\vec q}_{12} \right) \, 
\langle \xi \mid 
\left( {\vec \sigma}_1 \cdot {\vec q}_1 \right) \, 
\left( {\vec \sigma}_4 \cdot {\vec q}_4 \right) \, 
\left( {\vec q}_{1} \times {\vec q}_{12} \right) \cdot {\vec \sigma}_2 \, 
\left( {\vec \tau}_{1} \times {\vec \tau}_3 \right) \cdot {\vec \tau}_4 \, 
 \mid \xi \rangle  \nonumber \\
=2 \,\left( {\vec q}_4 \cdot {\vec q}_{12} \right) \, 
\left( {\vec q}_{1} \times {\vec q}_{12} \right) \cdot
\left( {\vec q}_4 \times {\vec q}_{1} \right) \, ,
\label{meVa4}
\end{eqnarray}
\begin{eqnarray}
\langle \xi \mid V^{a}_5 \mid \xi \rangle 
\equiv 
-\langle \xi \mid 
\left( {\vec \sigma}_1 \cdot {\vec q}_1 \right) \, 
\left( {\vec \sigma}_4 \cdot {\vec q}_4 \right) \, 
\left( {\vec q}_{1} \times {\vec q}_{12} \right) \cdot {\vec \sigma}_2 \, 
\left( {\vec q}_{12} \times {\vec q}_{4} \right) \cdot {\vec \sigma}_3 \, 
 {\vec \tau}_{1} \cdot {\vec \tau}_4 \, 
 \mid \xi \rangle  \nonumber \\
= {\vec q}_{1} \cdot \left[ \left( {\vec q}_{4} \times 
\left( {\vec q}_{1} \times {\vec q}_{12} \right)
\right) \times 
\left( {\vec q}_{12} \times {\vec q}_{4} \right)
 \right] 
\ + \
\left[ \left( {\vec q}_{12} \times {\vec q}_{4} \right) \cdot {\vec q}_{1}  \right] \,
\left[ \left( {\vec q}_{1} \times {\vec q}_{12} \right) \cdot {\vec q}_{4}  \right] \,
\label{meVa5}
\end{eqnarray}
\begin{eqnarray}
\langle \xi \mid V^{c}_1 \mid \xi \rangle 
\equiv 
\left( {\vec q}_4 \cdot {\vec q}_{12} \right) \, 
\langle \xi \mid 
\left( {\vec \sigma}_1 \cdot {\vec q}_1 \right) \, 
\left( {\vec \sigma}_4 \cdot {\vec q}_4 \right) \, 
\left( {\vec \tau}_{1} \cdot {\vec \tau}_4 \right) \, 
\left( {\vec \tau}_{2} \cdot {\vec \tau}_3 \right) \, 
 \mid \xi \rangle  \nonumber \\
=\left( {\vec q}_1 \cdot {\vec q}_{4} \right) \, 
\left( {\vec q}_{12} \cdot {\vec q}_{4} \right) \, ,
\label{meVc1}
\end{eqnarray}
\begin{eqnarray}
\langle \xi \mid V^{c}_2 \mid \xi \rangle 
\equiv 
-\left( {\vec q}_4 \cdot {\vec q}_{12} \right) \, 
\langle \xi \mid 
\left( {\vec \sigma}_1 \cdot {\vec q}_1 \right) \, 
\left( {\vec \sigma}_4 \cdot {\vec q}_4 \right) \, 
\left( {\vec \tau}_{1} \cdot {\vec \tau}_3 \right) \, 
\left( {\vec \tau}_{2} \cdot {\vec \tau}_4 \right) \, 
 \mid \xi \rangle  \nonumber \\
=3 \, \left( {\vec q}_1 \cdot {\vec q}_{4} \right) \, 
\left( {\vec q}_{12} \cdot {\vec q}_{4} \right) \, ,
\label{meVc2}
\end{eqnarray}
\begin{eqnarray}
\langle \xi \mid V^{c}_3 \mid \xi \rangle 
\equiv 
\langle \xi \mid 
\left( {\vec \sigma}_1 \cdot {\vec q}_1 \right) \, 
\left( {\vec \sigma}_4 \cdot {\vec q}_4 \right) \, 
\left( {\vec q}_{12} \times {\vec q}_4 \right) \cdot {\vec \sigma}_3 \, 
\left( {\vec \tau}_{1} \times {\vec \tau}_2 \right) \cdot {\vec \tau}_4  \,
 \mid \xi \rangle  \nonumber \\
=2 \, \left( {\vec q}_{12} \times {\vec q}_{4} \right) \cdot
\left( {\vec q}_{4} \times {\vec q}_{1} \right) \, ,
\label{meVc3}
\end{eqnarray}
\begin{eqnarray}
\langle \xi \mid V^{e}_1 \mid \xi \rangle 
\equiv 
\langle \xi \mid 
\left( {\vec \sigma}_1 \cdot {\vec q}_{34} \right) \, 
\left( {\vec \sigma}_2 \cdot {\vec q}_2 \right) \, 
\left( {\vec \sigma}_3 \cdot {\vec q}_3 \right) \, 
\left( {\vec \sigma}_4 \cdot {\vec q}_4 \right) \, 
{\vec \tau}_{1} \cdot {\vec \tau}_2  \,
{\vec \tau}_{3} \cdot {\vec \tau}_4  \,
 \mid \xi \rangle  \nonumber \\
=
\left( {\vec q}_{3} \times {\vec q}_{2} \right) \cdot
\left( {\vec q}_{34} \times {\vec q}_{4} \right) \, 
+
2\, \left( {\vec q}_{34} \times {\vec q}_{2} \right) \cdot
\left( {\vec q}_{3} \times {\vec q}_{4} \right) \, 
+
5\, \left( {\vec q}_{3} \cdot {\vec q}_{2} \right) \,
\left( {\vec q}_{34} \cdot {\vec q}_{4} \right) \, ,
\label{meVe1}
\end{eqnarray}
\begin{eqnarray}
\langle \xi \mid V^{f}_1 \mid \xi \rangle 
\equiv 
\langle \xi \mid 
\left( {\vec \sigma}_1 \cdot {\vec q}_{1} \right) \, 
\left( {\vec \sigma}_2 \cdot {\vec q}_2 \right) \, 
\left( {\vec \sigma}_3 \cdot {\vec q}_3 \right) \, 
\left( {\vec \sigma}_4 \cdot {\vec q}_4 \right) \, 
{\vec \tau}_{1} \cdot {\vec \tau}_2  \,
{\vec \tau}_{3} \cdot {\vec \tau}_4  \,
 \mid \xi \rangle  \nonumber \\
=
\left( {\vec q}_{3} \times {\vec q}_{1} \right) \cdot
\left( {\vec q}_{2} \times {\vec q}_{4} \right) \, 
+
2\, \left( {\vec q}_{2} \times {\vec q}_{1} \right) \cdot
\left( {\vec q}_{3} \times {\vec q}_{4} \right) \, 
+
5\, \left( {\vec q}_{3} \cdot {\vec q}_{1} \right) \,
\left( {\vec q}_{2} \cdot {\vec q}_{4} \right) \, ,
\label{meVf1}
\end{eqnarray}
\begin{eqnarray}
\langle \xi \mid V^{k}_1 \mid \xi \rangle 
\equiv 
\langle \xi \mid 
\left( {\vec \sigma}_1 \cdot {\vec q}_{1} \right) \, 
{\vec \sigma}_2 \cdot \left( {\vec q}_1 \times {\vec q}_{12} \right) \, 
\left( {\vec \sigma}_3 \times {\vec \sigma}_4 \right) \cdot {\vec q}_{12} \,
{\vec \tau}_{1} \cdot {\vec \tau}_3  \,
 \mid \xi \rangle  \nonumber \\
=
-2 \, \left( {\vec q}_{12} \times {\vec q}_{1} \right)^2 \, ,
\label{meVk1}
\end{eqnarray}
\begin{eqnarray}
\langle \xi \mid V^{k}_2 \mid \xi \rangle 
\equiv 
-\,\left( {\vec q}_{1} \cdot {\vec q}_{12} \right) \, 
\langle \xi \mid 
\left( {\vec \sigma}_1 \cdot {\vec q}_{1} \right) \, 
\left( {\vec \sigma}_3 \times {\vec \sigma}_4 \right) \cdot {\vec q}_{12} 
\left( {\vec \tau}_{1} \times {\vec \tau}_2 \right) \cdot {\vec \tau}_{3} \,
 \mid \xi \rangle  \nonumber \\
=
4 \, \left( {\vec q}_{1} \cdot {\vec q}_{12} \right)^2 \, ,
\label{meVk2}
\end{eqnarray}
\begin{eqnarray}
\langle \xi \mid V^{l}_1 \mid \xi \rangle 
\equiv 
\langle \xi \mid 
\left( {\vec \sigma}_1 \cdot {\vec q}_{1} \right) \, 
\left( {\vec \sigma}_3 \times {\vec \sigma}_4 \right) \cdot {\vec q}_{12} 
\left( {\vec \tau}_{1} \times {\vec \tau}_2 \right) \cdot {\vec \tau}_{3} \,
 \mid \xi \rangle  \nonumber \\
=
-4 \, \left( {\vec q}_{1} \cdot {\vec q}_{12} \right) \, ,
\label{meVl1}
\end{eqnarray}
\begin{eqnarray}
\langle \xi \mid V^{n}_1 \mid \xi \rangle 
\equiv 
\langle \xi \mid 
\left( {\vec \sigma}_1 \times {\vec \sigma}_2 \right) \cdot {\vec q}_{12}  \,
\left( {\vec \sigma}_3 \times {\vec \sigma}_4 \right) \cdot {\vec q}_{12} \,
{\vec \tau}_{2} \cdot {\vec \tau}_{3} \,
 \mid \xi \rangle  \nonumber \\
=
-4 \, {\vec q}_{12}^{\ 2} \, .
\label{meVn1}
\end{eqnarray}

Once the matrix elements (\ref{meVa1})--(\ref{meVn1}) have been calculated
we are left with the eighteen fold momentum space integrals. 
They can be written as
\begin{eqnarray}
\langle \Psi \mid V_{4N} \mid \Psi \rangle = \hfill \nonumber \\
\frac{24}{\left( 2 \pi \right)^9 } \,
\int \! d {\vec p} \,
\int \! d {\vec q} \,
\int \! d {\vec t} \,
\int \! d {\vec p}^{\,\prime} \,
\int \! d {\vec q}^{\,\prime} \,
\int \! d {\vec t}^{\,\prime} \,
g_{\Lambda_4} (x) \,
f(x)\,
V^i \left( {\vec p} , {\vec q}, {\vec t} , {\vec p}^{\,\prime},  {\vec q}^{\,\prime} , {\vec t}^{\,\prime} \right) \,
f(x^\prime) \,
g_{\Lambda_4} (x^\prime) \, ,
\label{integral18}
\end{eqnarray}
where the functions 
$ V^i \left( {\vec p} , {\vec q}, {\vec t} , {\vec p}^{\,\prime},  {\vec q}^{\,\prime} , {\vec t}^{\,\prime} \right) $
arise from introducing (\ref{indiv}) into  (\ref{meVa1})--(\ref{meVn1}) 
and the remaining expressions in (\ref{4nf}).
The additional factors $\frac1{\left( 2 \pi \right)^9 }$ and $24$ arise 
from the wave function normalization and due to the fact that all nucleons' permutations 
yield the same result in the case of the totally antisymmetric wave function.
The functions $g_{\Lambda_4} (x)$ and $g_{\Lambda_4} (x^\prime)$ are introduced since the 
expressions in (\ref{4nf}) need to be regularized. We choose a simple form
\begin{eqnarray}
g_{\Lambda_4} (x) \, = \, \exp\left[ - \left( \frac{m x}{2 \Lambda_4^2 }\right)^3 \right] ,
\label{g}
\end{eqnarray}
so all the results will depend on the parameter $\Lambda_4$.

\begin{figure}[hp]\centering
\epsfig{file=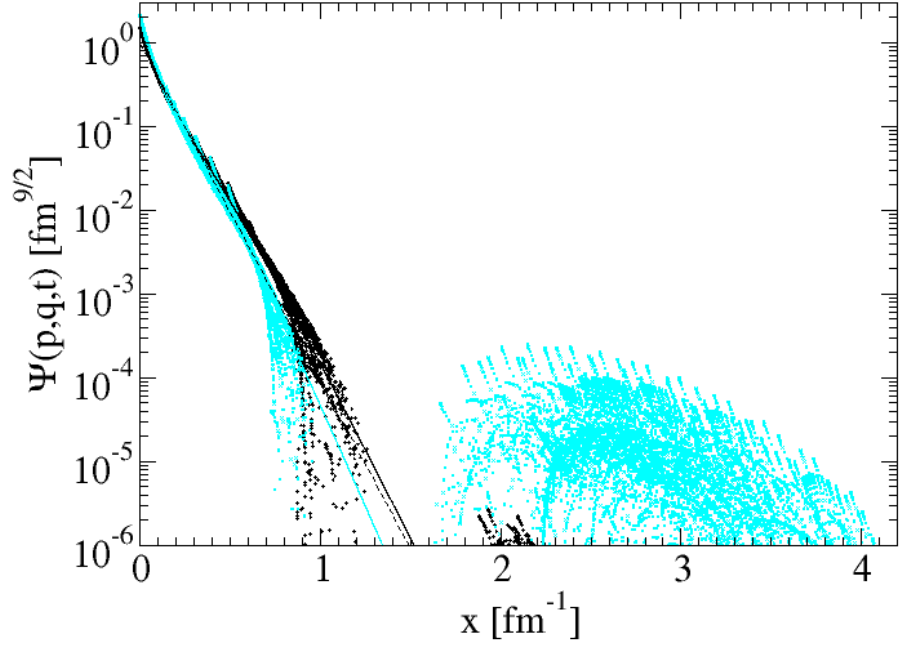,width=12cm,angle=0}
\caption{
The values of $\Psi_0 (p, q, t )$ for all possible combinations 
of $p$, $q$ and $t$ are plotted as a function of
$x = \frac1{m} \left( {p}^{2} + \frac34 {q}^{2}  + \frac23 {t}^{2} \right) $ 
for the $(\Lambda, \tilde{\Lambda})$= $(400 \, {\rm MeV/c}, 500 \, {\rm MeV/c})$
case with black symbols 
and for the $(\Lambda, \tilde{\Lambda})$= $(550 \, {\rm MeV/c}, 500 \, {\rm MeV/c})$
case with grey (cyan in color) points.
The corresponding fits are represented by lines of the same color.
The dashed line shows the Gaussian wave function (\ref{f1}).
}
\label{fig2}
\end{figure}

We consider two types of the $^4$He wave functions. 
First one is a pure model Gaussian function \cite{thompson69}
\begin{eqnarray}
f_1 (x) \, = \, \frac{2^{3/2}}{\beta^{9/4} \, \pi^{9/4}} \, \exp\left( - \frac{m}{\beta} \, x \right) ,
\label{f1}
\end{eqnarray}
where the value of the parameter $\beta$ is chosen after \cite{thompson69} as 0.514 fm$^{-2}$.
Wave functions of the second type are obtained in a quite different manner. 
We consider the wave functions which are solutions of the Schr\"odinger
equation with the NLO chiral potentials \cite{NLO,SFR}
labeled by the following sets of the parameters $(\Lambda, \tilde{\Lambda})$:
$(400 \, {\rm MeV/c}, 500 \, {\rm MeV/c})$,
$(550 \, {\rm MeV/c}, 500 \, {\rm MeV/c})$,
$(550 \, {\rm MeV/c}, 600 \, {\rm MeV/c})$,
$(400 \, {\rm MeV/c}, 700 \, {\rm MeV/c})$ and
$(550 \, {\rm MeV/c}, 700 \, {\rm MeV/c})$.
We checked that the wave functions with the same parameter ${\Lambda}$
have very similar properties so the dependence on $\tilde{\Lambda}$
is very weak. Thus we restricted ourselves to two cases only: 
$(\Lambda, \tilde{\Lambda})$=
$(400 \, {\rm MeV/c}, 500 \, {\rm MeV/c})$ and
$(550 \, {\rm MeV/c}, 500 \, {\rm MeV/c})$.
For these two wave functions gained 
by rigorous solutions of the 4N Faddeev-Yakubovsky equations
we extracted the component with the totally 
antisymmetric spin-isospin part. 
In both cases this component is dominant. It constitutes 94.3 \% (88.7 \%)
of the original $(400 \, {\rm MeV/c}, 500 \, {\rm MeV/c})$
($(550 \, {\rm MeV/c}, 500 \, {\rm MeV/c})$) wave function.  
Further we removed all contributions from the states
with non-zero angular momenta. These components are small 
and represent only 0.3 \% and 2.9 \% of the corresponding full wave functions.
In this way we end up with the wave function components depending only on magnitudes 
of the momenta ${\vec p}$, ${\vec q}$ and ${\vec t}$,
$\Psi_0 ( p, q, t )$, given on a certain grid.
In order to facilitate the calculations, we represented $\Psi_0 ( p, q, t )$
by a one variable formula analogous to (\ref{f1}):
\begin{eqnarray}
f_2 (x) \, = 
\left( a_0 + a_1 x^{a_2} \right) \, \exp\left( -a_3 x^{a_4} \right) \, ,
\label{f2}
\end{eqnarray}
with the parameters $a_0$, $a_1$, $a_2$, $a_3$ and $a_4$ given in Table~I.
\begin{table}[tp]
\begin{center}
\begin{tabular}[t]{ccccccc}
\hline
wave function & $a_0$  & $a_1$ &  $a_2$ &  $a_3$ & $a_4$ \\
\hline
$(400 \, {\rm MeV/c}, 500 \, {\rm MeV/c})$  & 1.53266 & 40.4324 & 2.36626 & 12.5715 & 0.927233 \\
$(550 \, {\rm MeV/c}, 500 \, {\rm MeV/c})$  & 2.12619 & 86.6989 & 2.41787 & 14.4705 & 0.921551 \\
\end{tabular}
\caption{
Parameters of the one dimensional fits (\ref{f2}) for the two
chiral NLO wave functions considered in this paper.
}
\end{center}
\label{tab1}
\end{table}
For the reader's orientation we show in Fig.~\ref{fig2} 
the components $\Psi_0 ( p, q, t )$ of the two chiral wave functions
plotted as a function of 
$x = \frac1{m} \left( {p}^{2} + \frac34 {q}^{2}  + \frac23 {t}^{2} \right) $
together with the lines fitted according to (\ref{f2}).
It is clear that the fits can be considered to be reasonable approximations 
to the underlying  $\Psi_0 (p, q, t )$ components only for small values of $x$.
At larger $x$ the values of $\Psi_0 (p, q, t )$ are clearly underestimated.
However, we assume in this first attempt that the main contributions
to the expectation values come from the $x$ region,
for which the fits still reflect the bulk properties of the original  
$\Psi_0 ( p, q, t )$. That is why in the actual calculations we could use
the simple analytical forms of  (\ref{f2}).
Note that the very simple Gaussian wave function is close 
to the NLO fit with $\Lambda$= 400 MeV/c.

In the practical calculations we used
the basic Monte Carlo method and generated uniform distributions
in each of the eighteen dimensions by means of the portable random number 
generator {\em ran2} from Ref.~\cite{nr}. 
We found it sufficient to restrict the 
magnitudes of the relative momenta $p$, $q$ and $t$ to the 
following values $p_{\rm max}$= $q_{\rm max}$= $t_{\rm max}$= 6 fm$^{-1}$.
As primary tests of our Monte Carlo calculations we checked 
the norm and the internal kinetic energy of $^4$He. 
These quantities can be calculated very precisely as three fold integrals
but for tests were written as nine fold and (squared) even as eighteen 
fold integrals. The 4N force expectation values are approximated by
\begin{eqnarray}
I \equiv \int f dv \approx \frac{v}{N} \, \sum\limits_{i=1}^N f (x_i) ,
\label{montecarlo}
\end{eqnarray}
for which the one standard deviation error estimate reads
\begin{eqnarray}
\delta(I) = 
\frac{v}{\sqrt{N}} \, \sqrt{   \sum\limits_{i=1}^N f^2 (x_i) \, - \, \left(  \sum\limits_{i=1}^N f (x_i) \right)^2 } .
\label{montecarlo2}
\end{eqnarray}
Here the points $x_1, x_2, \dots, x_N$ are uniformly distributed in the eighteen dimensional volume $v$.
Tables~II and III show our results for the Gaussian function $f_1 (x)$
and the two first chiral NLO wave functions from Tab.~I.
We used $N$= 10$^9$ integral points.
\begin{table}[tp]
\begin{center}
\begin{tabular}[t]{c|c|c}
\hline
parts & $I$  & $\delta(I)$  \\
of the 4N force                      & MeV  & MeV  \\
\hline
$V^a$  & -0.002906 & 22 $\times$ 10$^{-6}$  \\
$V^c$  & -0.005557 & 25 $\times$ 10$^{-6}$  \\
$V^e$  & -0.008462 & 20 $\times$ 10$^{-6}$  \\
$V^f$  &  0.005692 & 12 $\times$ 10$^{-6}$  \\
$V^k$  &  0.0005925 $C_T$ & 19 $\times$ 10$^{-7}$ $C_T$ \\
$V^l$  & 0.000622657 $C_T$ & 10 $\times$ 10$^{-7}$ $C_T$ \\
$V^n$  & -0.000046044 $C_T^2$ & 55 $\times$ 10$^{-9}$ $C_T^2$ \\
\end{tabular}
\caption{
Expectation values of the individual parts of the 4N force
for the Gaussian wave function $f_1(x)$. The regulator function
$g_{\Lambda_4} (x)$ with $\Lambda_4$= 500 MeV/c is used. For the three last terms
the value of the low energy constant (LEC) $C_T$ in GeV$^{-2}$ 
should be inserted. All the numbers should be additionally multiplied by the factor 24.
}
\end{center}
\label{tab2}
\end{table}

\begin{table}[tp]
\begin{center}
\begin{tabular}[t]{c|c|c|c|c}
\hline
parts  & $I (400,500)$  & $\delta(I) (400,500)$  & $I (550,500)$  & $\delta(I) (550,500)$ \\
of the 4N force                     & MeV  & MeV & MeV & MeV  \\
\hline
$V^a$  & -0.00434503           & 17 $\times$ 10$^{-5}$         & -0.00222788  & 10 $\times$ 10$^{-5}$ \\
$V^c$  & -0.0084033            & 19 $\times$ 10$^{-5}$         & -0.00445691  & 12 $\times$ 10$^{-5}$ \\
$V^e$  & -0.0133568            & 15 $\times$ 10$^{-5}$         & -0.00683624  & 92 $\times$ 10$^{-6}$ \\
$V^f$  &  0.00914028           & 93 $\times$ 10$^{-6}$         &  0.00460722  & 56 $\times$ 10$^{-6}$ \\
$V^k$  &  0.000926931 $C_T$    & 14 $\times$ 10$^{-6}$ $C_T$   &  0.000489232 $C_T$ & 86 $\times$ 10$^{-7}$ $C_T$ \\
$V^l$  &  0.000964454 $C_T$    & 77 $\times$ 10$^{-7}$ $C_T$   &  0.000512526 $C_T$ & 49 $\times$ 10$^{-7}$ $C_T$ \\
$V^n$  &  -0.00007243 $C_T^2 $ & 41 $\times$ 10$^{-8}$ $C_T^2$ & -0.00003812 $C_T^2$ & 26 $\times$ 10$^{-8}$ $C_T^2$ \\
\end{tabular}
\caption{
The same as in Tab.~I for the two chiral NLO wave functions.
Note that all the values should be additionally
corrected for the norms of the wave functions:
$\langle \Psi \mid \Psi \rangle$= 1.093 
($(\Lambda, \tilde{\Lambda})$= $(400 \, {\rm MeV/c}, 500 \, {\rm MeV/c})$)
and $\langle \Psi \mid \Psi \rangle$=1.011 
($(\Lambda, \tilde{\Lambda})$= $(550 \, {\rm MeV/c}, 500 \, {\rm MeV/c})$).
As in Tab.~I, the factor of 24 is not included.
}
\end{center}
\label{tab4}
\end{table}

\begin{figure}[tb]
\includegraphics[width=8.5cm,keepaspectratio,angle=-90,clip]{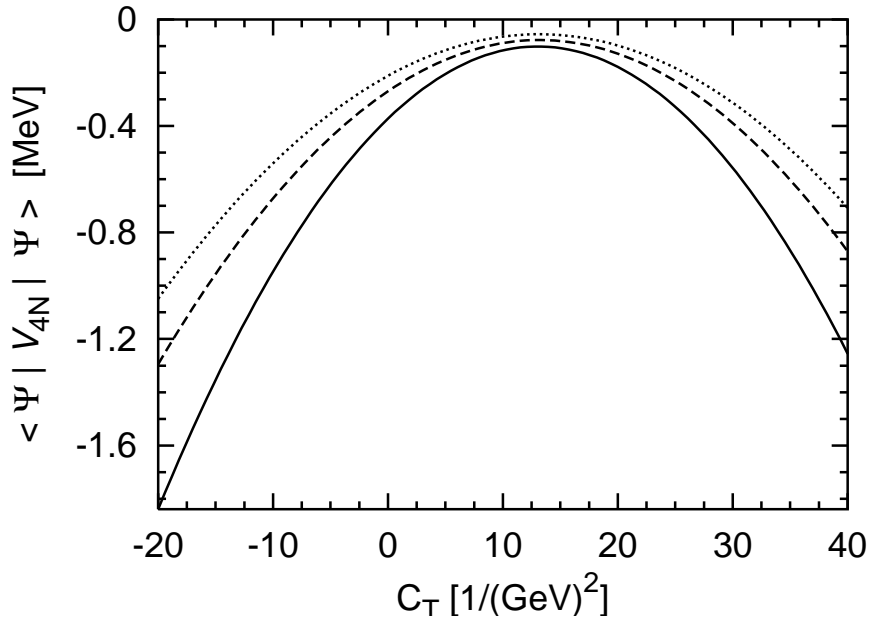}
\caption{
The expectation values of the 4N force for $\Lambda_4$= 500 MeV/c
as a function of the $C_T$ LEC for different parametrizations of the $^4$He wave function.
The dashed line represents the Gaussian wave function (\ref{f1}),
the dotted line corresponds to the case
of $(\Lambda, \tilde{\Lambda})$= $(550 \, {\rm MeV/c}, 500 \, {\rm MeV/c})$
and the solid line is for 
$(\Lambda, \tilde{\Lambda})$= $(400 \, {\rm MeV/c}, 500 \, {\rm MeV/c})$.
\label{fig3}
}
\end{figure}

We show also in Fig.~\ref{fig3} the expectation values of the 4N force 
as a function of the $C_T$ LEC. This is our final prediction, which includes all 
the required corrections.
For $C_T \approx$ 13 GeV$^{-2}$ the magnitudes of the sum of the
expectation values reach their minimum
and we obtain approximately -0.077, -0.107 and -0.061 MeV
for the three wave functions (Gaussian, $(\Lambda, \tilde{\Lambda})$= 
$(400 \, {\rm MeV/c}, 500 \, {\rm MeV/c})$, 
$(\Lambda, \tilde{\Lambda})$= $(550 \, {\rm MeV/c}, 500 \, {\rm MeV/c})$)
considered, respectively.
For $C_T$=0 the corresponding numbers are
-0.270, -0.386 and -0.219 MeV.
Only the two parametrizations of the chiral wave functions
are consistent, at least to some extent, with the 4N potential.
Thus we can state that the 4N force effects might vary from a few tens of keV 
to 1-2 MeV. Note that not the whole range of the $C_T$ values 
shown in the figures actually appears for different orders 
of the chiral expansion~\cite{SFR}.

\begin{figure}[tb]
\includegraphics[width=8.5cm,keepaspectratio,angle=-90,clip]{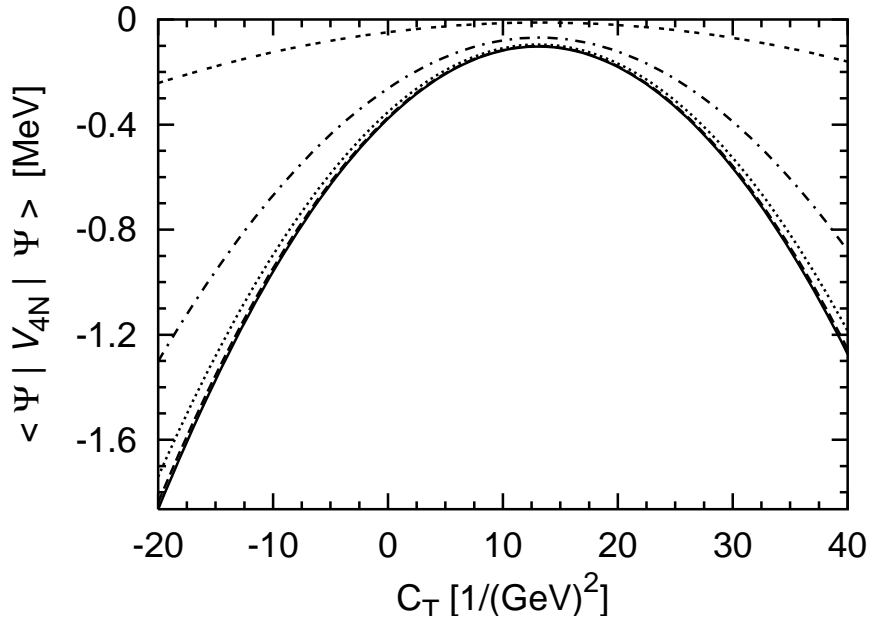}
\caption{
The expectation values of the 4N force
as a function of the $C_T$ LEC for the first chiral 
wave function ($(\Lambda, \tilde{\Lambda})$= $(400 \, {\rm MeV/c}, 500 \, {\rm MeV/c})$)
calculated with different parameters $\Lambda_4$:
200 MeV/c (double dashed line), 300 MeV/c (dash-dotted line),
400 MeV/c (dotted line), 500 MeV/c (dashed line) and 600 MeV/c (solid line).
\label{fig4}
}
\end{figure}

It remains to check the influence of different regulator functions 
on our predictions. To this aim we took the first chiral wave function and 
calculated the expectation values additionally with $\Lambda_4$= 200, 300, 400 and 600 MeV/c.
As can be seen in Fig.~\ref{fig4}, the results do not differ much from each other
for $\Lambda_4 >$ 300 MeV/c. Note that our definition of the regulator function 
$g_{\Lambda_4} (x) $ given in (\ref{g}) introduces an additional factor of 2, 
as compared for example with \cite{SFR,CUT}. Thus the values 300 and 400 MeV/c for
$\Lambda_4$ multiplied 
by $\sqrt{2}$ roughly correspond to the parameters $\Lambda$ (400 - 550 MeV/c)
of the wave functions.

\section{Summary}
\label{sec:4}

We estimated for the first time 4N force effects in $^4$He
by calculating explicitly the expectations values 
of different 4N force parts between several $^4$He wave functions. 
Our estimates agree qualitatively with 
modern nuclear force predictions for the $\alpha$ particle \cite{INDIRECT},
which do not leave much room for the action of 4N forces.
Our predictions lack full consistency between 
the wave functions and the 4N potential and also
neglect smaller components of $^4$He.
The strong dependence of the expectation value on $C_T$ 
in the considered interval will probably be reduced using 
a fully consistent $^4$He wave function at order NNNLO.
Nevertheless, our results give some hint how important 4N 
force effects might be.

\acknowledgments
This work was supported by the Polish Committee for Scientific Research
under grant no. 2P03B00825, by the NATO grant no. PST.CLG.978943,
and by DOE under grants nos. DE-FG03-00ER41132 and DE-FC02-01ER41187,
and by the Helmholtz Association, contract number VH-NG-222. 
One of the authors (EE) acknowledges financial support from the Thomas Jefferson 
National Accelerator Facility, USA.



\begin{thebibliography}{9}

\bibitem{4nfpaper} E. Epelbaum, nucl-th/0511025.

\bibitem{thompson69} D.R. Thompson, I. Reichstein, W. McClure, and Y. C. Tang, 
Phys. Rev. {\bf 185}, 1351 (1969).

\bibitem{NLO} E. Epelbaum, W. Gl\"ockle, Ulf.-G. Mei{\ss}ner,
Nucl. Phys. {\bf A747}, 362 (2005).

\bibitem{SFR} E. Epelbaum, nucl-th/0509032.

\bibitem{nr} W. H. Press, B. P. Flannery, S. A. Teukolsky, W. T. Vetterling,
{\em Numerical Recipes in C: The Art of Scientific Computing},
Cambridge University Press, 1988.

\bibitem{CUT} E. Epelbaum, A. Nogga, W. Gl\"ockle, H. Kamada, Ulf.-G. Mei{\ss}ner, H. Wita{\l}a,
Phys. Rev. C{\bf 66}, 064001 (2002).

\bibitem{INDIRECT} A. Nogga, H. Kamada, and W. Gl\"ockle,
Phys. Rev. Lett. {\bf 85}, 944 (2000).

\end{thebibliography}
\end{document}